\documentclass[english]{article}
\usepackage[T1]{fontenc}
\usepackage[latin9]{inputenc}
\usepackage{color}
\usepackage{amssymb}
\usepackage{babel}
\begin{document}

\title{\textbf{Bohr-like black holes}}

\author{\textbf{Christian Corda}}

\maketitle
\begin{center}
Dipartimento di Fisica e Chimica, Istituto Universitario di Ricerca
Scientifica \textquotedbl{}Santa Rita\textquotedbl{}, Centro di Scienze
Naturali, Via di Galceti, 74, 59100 Prato
\par\end{center}

\begin{center}
Institute for Theoretical Physics and Advanced Mathematics (IFM) Einstein-Galilei,
Via Santa Gonda 14, 59100 Prato, Italy
\par\end{center}

\begin{center}
International Institute for Applicable Mathematics \& Information
Sciences (IIAMIS),  B.M. Birla Science Centre, Adarsh Nagar, Hyderabad
- 500 463, India 
\par\end{center}

\begin{center}
\textit{E-mail address:} \textcolor{blue}{cordac.galilei@gmail.com} 
\par\end{center}
\begin{abstract}
The idea that black holes (BHs) result in highly excited states representing
both the ``hydrogen atom'' and the ``quasi-thermal emission''
in quantum gravity is today an intuitive but general conviction. In
this paper it will be shown that such an intuitive picture is more
than a picture. In fact, we will discuss a model of quantum BH somewhat
similar to the historical semi-classical model of the structure of
a hydrogen atom introduced by Bohr in 1913. The model is completely
consistent with existing results in the literature, starting from
the celebrated result of Bekenstein on the area quantization. 
\end{abstract}
Researchers in quantum gravity {[}20{]} intuitively think that, in
some respects, BHs are the fundamental bricks of quantum gravity in
the same way that atoms are the fundamental bricks of quantum mechanics.
This analogy suggests that the BH mass should have a discrete spectrum.
In this extended abstract, we show that the such an intuitive picture
is more than a picture. Starting from the natural correspondence between
Hawking radiation {[}1{]} and BH quasi-normal modes (QNMs) {[}2\textendash{}4{]},
we show that QNMs can be really interpreted in terms of BH quantum
levels discussing a BH model somewhat similar to the semi-classical
Bohr model of the structure of a hydrogen atom {[}5, 6{]}. 

One considers Dirac delta perturbations {[}2\textendash{}4, 7{]} representing
subsequent absorptions of particles having negative energies which
are associated to emissions of Hawking quanta in the mechanism of
particle pair creation. BH responses to perturbations are QNMs {[}2\textendash{}4,
8-12, 21{]}, which are frequencies of radial spin-$j$ perturbations
obeying a time independent Schröedinger-like equation {[}2\textendash{}4,
12{]}. They are the BH modes of energy dissipation which frequency
is allowed to be complex {[}2\textendash{}4, 12{]}. For large values
of the principal quantum number $n$, where $n=1,2,...$, QNMs become
independent of both the spin and the angular momentum quantum numbers
{[}2\textendash{}4, 8, 12, 13, 14{]}, in perfect agreement with \emph{Bohr's
Correspondence Principle} {[}15{]}, which states that \textquotedblleft{}transition
frequencies at large quantum numbers should equal classical oscillation
frequencies\textquotedblright{}. In other words, Bohr's Correspondence
Principle enables an accurate semi-classical analysis for large values
of the principal quantum number $n,$ i.e, for excited BHs. By using
that principle, Hod has shown that QNMs release information about
the area quantization as QNMs are associated to absorption of particles
{[}13, 44{]}. Hod's work was refined by Maggiore {[}8{]} who solved
some important problems. On the other hand, as QNMs are \emph{countable}
frequencies, ideas on the \emph{continuous} character of Hawking radiation
did not agree with attempts to interpret QNMs in terms of emitted
quanta, preventing to associate QNMs to Hawking radiation {[}12{]}.
Recently, ourselves and collaborators {[}2\textendash{}4, 8-11, 21{]}
observed that the non-thermal spectrum of Parikh and Wilczek {[}16{]}
also implies the countable character of subsequent emissions of Hawking
quanta. This issue enables a natural correspondence between QNMs and
Hawking radiation, permitting to interpret QNMs also in terms of emitted
energies {[}2\textendash{}4, 8-11{]}. In fact, Dirac delta perturbations
due to discrete subsequent absorptions of particles having negative
energies, which are associated to emissions of Hawking quanta in the
mechanism of particle pair creation by quantum fluctuations, generates
BH QNMs {[}2\textendash{}4, 8-11{]}. On the other hand, the correspondence
between emitted radiation and proper oscillation of the emitting body
is a fundamental behavior of every radiation process in science. Based
on such a natural correspondence between Hawking radiation and BH
QNMs, one can consider QNMs in terms of quantum levels also for emitted
energies {[}2\textendash{}4, 8-11{]}. For large values of the principal
quantum number $n,$ i.e, for excited BHs, and independently of the
angular momentum quantum number, the QNMs expression of the Schwarzschild
BH which takes into account the non-strictly thermal behavior of the
radiation spectrum is obtained as {[}2\textendash{}4{]}

\noindent 
\begin{equation}
\omega_{n}=a+ib+\frac{in}{4M-2|\omega_{n}|}\backsimeq\frac{in}{4M-2|\omega_{n}|},\label{eq: quasinormal modes corrected}
\end{equation}
where $a$ and $b$ are real numbers with $a=\frac{\ln3}{4\pi(2M-|\omega_{n}|)},\; b=\frac{1}{4(2M-|\omega_{n}|)}$
for $j=0,2$ (scalar and gravitational perturbations), $a=0,\; b=0$
for $j=1$ (vector perturbations) and $a=0,\; b=\frac{1}{4(2M-|\omega_{n}|)}$
for half-integer values of $j$. On the other hand, as $a,b\ll|\frac{in}{4M-2|\omega_{n}|}|$,
a fundamental consequence is that the quantum of area obtained from
the asymptotic values of $|\omega_{n}|$ is an intrinsic property
of Schwarzschild BHs because for large $n$ the leading asymptotic
behavior of $|\omega_{n}|$ is given by the leading term in the imaginary
part of the complex frequencies, and it does not depend on the spin
content of the perturbation {[}2\textendash{}4, 8{]}. An intuitive
derivation of eq. (\ref{eq: quasinormal modes corrected}) can be
found in {[}3, 4{]}. We \emph{rigorously} derived such an equation
in the Appendix of {[}2{]}. If one considers the leading asymptotic
behavior, the physical solution for the absolute values of the frequencies
(\ref{eq: quasinormal modes corrected}) is {[}2\textendash{}4{]}

\noindent 
\begin{equation}
E_{n}\equiv|\omega_{n}|=M-\sqrt{M^{2}-\frac{n}{2}}.\label{eq: radice fisica}
\end{equation}
$E_{n}\:$ is interpreted like the total energy emitted by the BH
at that time, i.e. when the BH is excited at a level $n$ {[}2\textendash{}4{]}.
Considering an emission from the ground state (i.e. a BH which is
not excited) to a state with large $n=n_{1}$ and using eq. (\ref{eq: radice fisica}),
the BH mass changes from $M\:$ to {[}2\textendash{}4{]}

\begin{equation}
M_{n_{1}}\equiv M-E_{n_{1}}=\sqrt{M^{2}-\frac{n_{1}}{2}}.\label{eq: me-1}
\end{equation}
In the transition from the state with $n=n_{1}$ to a state with $n=n_{2}$
where $n_{2}>n_{1}$ the BH mass changes again from $M_{n_{1}}\:$
to

\begin{equation}
\begin{array}{c}
M_{n_{2}}\equiv M_{n_{1}}-\Delta E_{n_{1}\rightarrow n_{2}}=M-E_{n_{2}}\\
=\sqrt{M^{2}-\frac{n_{2}}{2}},
\end{array}\label{eq: me}
\end{equation}
where 
\begin{equation}
\Delta E_{n_{1}\rightarrow n_{2}}\equiv E_{n_{2}}-E_{n_{1}}=M_{n_{1}}-M_{n_{2}}=\sqrt{M^{2}-\frac{n_{1}}{2}}-\sqrt{M^{2}-\frac{n_{2}}{2}},\label{eq: jump}
\end{equation}
is the jump between the two levels due to the emission of a particle
having frequency $\Delta E_{n_{1}\rightarrow n_{2}}$. Thus, in our
BH model {[}2{]}, during a quantum jump a discrete amount of energy
is radiated and, for large values of the principal quantum number
$n,$ the analysis becomes independent of the other quantum numbers.
In a certain sense, QNMs represent the \textquotedbl{}electron\textquotedbl{}
which jumps from a level to another one and the absolute values of
the QNMs frequencies represent the energy \textquotedbl{}shells\textquotedbl{}
{[}2{]}. In Bohr model {[}5, 6{]} electrons can only gain and lose
energy by jumping from one allowed energy shell to another, absorbing
or emitting radiation with an energy difference of the levels according
to the Planck relation (in standard units) $E=hf$, where $\: h\:$
is the Planck constant and $f\:$ the transition frequency. In our
BH model {[}2{]}, QNMs can only gain and lose energy by jumping from
one allowed energy shell to another, absorbing or emitting radiation
(emitted radiation is given by Hawking quanta) with an energy difference
of the levels according to eq. (\ref{eq: jump}). The similarity is
completed if one notes that the interpretation of eq. (\ref{eq: radice fisica})
is of a particle, the ``electron'', quantized on a circle of length
{[}3{]} 
\begin{equation}
L=\frac{1}{T_{E}(E_{n})}=4\pi\left(M+\sqrt{M^{2}-\frac{n}{2}}\right),\label{eq: lunghezza cerchio}
\end{equation}
which is the analogous of the electron travelling in circular orbits
around the hydrogen nucleus, similar in structure to the solar system,
of Bohr model {[}5, 6{]}. On the other hand, Bohr model is an approximated
model of the hydrogen atom with respect to the valence shell atom
model of full quantum mechanics. In the same way, our BH model should
be an approximated model with respect to the definitive, but at the
present time unknown, BH model arising from a full quantum gravity
theory. 

Let us discuss \emph{the area quantization}. Setting $n_{1}=n-1$,
$n_{2}=n$ in eq. (\ref{eq: jump}) on gets the emitted energy for
a jump among two neighboring levels {[}2, 3, 4{]}

\noindent 
\begin{equation}
\Delta E_{n-1\rightarrow n}=\sqrt{M^{2}-\frac{n+1}{2}}-\sqrt{M^{2}-\frac{n}{2}}.\label{eq: variazione}
\end{equation}
An enlightening analysis that we rigorously developed in {[}2{]} shows
that eq. (\ref{eq: variazione}) leads to the area quantum

\begin{equation}
|\triangle A_{n}|=|\triangle A_{n-1}|=8\pi,\label{eq: 8 pi planck}
\end{equation}
which is exactly the famous result of Bekenstein on the area quantization
{[}17{]}, and this \emph{cannot} be a coincidence. Other fundamental
results are: i) the famous formula of Bekenstein-Hawking entropy {[}1,
18, 19{]} reads {[}2{]} 
\begin{equation}
\left(S_{BH}\right)_{n-1}\equiv\frac{A_{n-1}}{4}=8\pi N_{n-1}M_{n-1}\cdot\Delta E_{n-1\rightarrow n}=4\pi\left(M^{2}-\frac{n-1}{2}\right)\label{eq: Bekenstein-Hawking  n-1}
\end{equation}
before the emission and 
\begin{equation}
\left(S_{BH}\right)_{n}\equiv\frac{A_{n}}{4}=8\pi N_{n}M_{n}\cdot\Delta E_{n-1\rightarrow n}=4\pi\left(M^{2}-\frac{n}{2}\right),\label{eq: Bekenstein-Hawking  n}
\end{equation}
after the emission, respectively; ii) the total BH entropy becomes
{[}2{]}

\noindent 
\begin{equation}
\begin{array}{c}
\left(S_{total}\right)_{n-1}=4\pi\left(M^{2}-\frac{n-1}{2}\right)\\
\\
-\ln\left[4\pi\left(M^{2}-\frac{n-1}{2}\right)\right]+\frac{3}{32\pi\left(M^{2}-\frac{n-1}{2}\right))}
\end{array}\label{eq: entropia n-1}
\end{equation}

\noindent before the emission, and 
\begin{equation}
\begin{array}{c}
\left(S_{total}\right)_{n}=4\pi\left(M^{2}-\frac{n}{2}\right)\\
\\
-\ln\left[4\pi\left(M^{2}-\frac{n}{2}\right)\right]+\frac{3}{32\pi\left(M^{2}-\frac{n}{2}\right)}
\end{array}\label{eq: entropia n}
\end{equation}
after the emission, respectively. Thus, both the Bekenstein-Hawking
entropy and the total BH entropy results a function of the BH excited
state $n.$ We stress that our results are in perfect agreement with
existing results in the literature, see {[}2-4{]} for details.

\textbf{Conclusion remarks}

We have shown that the intuitive but general conviction that BHs result
in highly excited states representing both the ``hydrogen atom''
and the ``quasi-thermal emission'' in quantum gravity is more than
a picture as we have indeed discussed a model of quantum BH somewhat
similar to the historical semi-classical model of the structure of
a hydrogen atom introduced by Bohr in 1913. This Bohr-like model of
BHs is totally consistent with existing results in the literature,
starting from the famous result of Bekenstein on the area quantization.
The issue that semi-classical BHs are the analogous of the Bohr model
for the hydrogen atom is also an intriguing starting point for future
work on a new approach to quantum gravity based on the ``electron''
represented by QNMs, i.e. for the potential construction of a ``QNMs
quantum gravity''.

\paragraph*{Acknowledgements}

It is a pleasure to thank Prof. T. Simos for inviting me to release
a Plenary Lecture in the 12th International Conference of Numerical
Analysis and Applied Mathematics. This Proceeding paper is indeed
an extended abstract of such a Plenary Lecture. I thank the referees
for useful comments and advices that permitted to improve this paper.

\subsubsection*{References}

{[}1{]} S. W. Hawking, Commun. Math. Phys. 43, 199 (1975).

{[}2{]} C. Corda, Eur. Phys. J. C 73, 2665 (2013). 

{[}3{]} C. Corda, Int. Journ. Mod. Phys. D 21, 1242023 (2012). 

{[}4{]} C. Corda, J. High En. Phys. 1108, 101 (2011).

{[}5{]} N. Bohr, Philos. Mag. 26 , 1 (1913).

{[}6{]} N. Bohr, Philos. Mag. 26 , 476 (1913).

{[}7{]} C. Corda, Ann. Phys. 337, 49 (2013). 

{[}8{]} M. Maggiore, Phys. Rev. Lett. 100, 141301 (2008). 

{[}9{]} C. Corda, Electr. Jour. Theor. Phys. 11, 30 27 (2014). 

{[}10{]} C. Corda, S. H. Hendi, R. Katebi, N. O. Schmidt, JHEP 06,
008 (2013). 

{[}11{]} C. Corda, S. H. Hendi, R. Katebi, N. O. Schmidt, Adv. High
En. Phys. 527874 (2014). 

{[}12{]} L. Motl, Adv. Theor. Math. Phys. 6, 1135 (2003).

{[}13{]} S. Hod, Gen. Rel. Grav. 31, 1639 (1999).

{[}14{]} S. Hod, Phys. Rev. Lett. 81 4293 (1998). 

{[}15{]} N. Bohr, Zeits. Phys. 2, 423 (1920). 

{[}16{]} M. K. Parikh and F. Wilczek, Phys. Rev. Lett. 85, 5042 (2000). 

{[}17{]} J. D. Bekenstein, Lett. Nuovo Cim. 11, 467 (1974). 

{[}18{]} J. D. Bekenstein, Nuovo Cim. Lett. 4, 737 (1972). 

{[}19{]} J. D. Bekenstein, Phys. Rev. D7, 2333 (1973). 

{[}20{]} C. Corda, New Adv. Phys., 7, 1, 67 (2013) 

{[}21{]} C. Corda, S. H. Hendi, R. Katebi, N. O. Schmidt, Adv. High
En. Phys. 530547 (2014).
\end{document}